\documentclass[conference]{IEEEtran}
\usepackage{graphicx} 
\usepackage{amsmath,amssymb}
\usepackage{xcolor}
\usepackage{bm}
\usepackage{algorithm}
\usepackage{algorithmic}
\usepackage{hyperref}

\def\BibTeX{{\rm B\kern-.05em{\sc i\kern-.025em b}\kern-.08em
    T\kern-.1667em\lower.7ex\hbox{E}\kern-.125emX}}

\title{SAGIPS: A Scalable Asynchronous Generative Inverse Problem Solver}

 \author{\IEEEauthorblockN{Daniel Lersch}
 \IEEEauthorblockA{\textit{Jefferson Lab Data Science Department} \\
 \textit{Thomas Jefferson National Accelerator Facility}\\
 Newport News, USA \\
 dlersch@jlab.org}
 \and
 \IEEEauthorblockN{Malachi Schram}
 \IEEEauthorblockA{\textit{Jefferson Lab Data Science Department} \\
 \textit{Thomas Jefferson National Accelerator Facility}\\
 Newport News, USA \\
 schram@jlab.org}
 \and
 \IEEEauthorblockN{Zhenyu Dai}
 \IEEEauthorblockA{\textit{Jefferson Lab Data Science Department} \\
 \textit{Thomas Jefferson National Accelerator Facility}\\
 Newport News, USA \\
 dai@jlab.org}
 \and
 \IEEEauthorblockN{Kishansingh Rajput}
 \IEEEauthorblockA{\textit{Jefferson Lab Data Science Department} \\
 \textit{Thomas Jefferson National Accelerator Facility}\\
 Newport News, USA \\
 kishan@jlab.org}
 \and
 \IEEEauthorblockN{Xingfu Wu}
 \IEEEauthorblockA{\textit{Mathematics and Computer Science Division} \\
 \textit{Argonne National Laboratory}\\
 Chicago, USA \\
 xingfu.wu@anl.gov}
 \and
 \IEEEauthorblockN{J. Taylor Childers}
 \IEEEauthorblockA{\textit{Leadership Computing Division} \\
 \textit{Argonne National Laboratory}\\
 Chicago, USA \\
 jchilders@anl.gov}
 }
\author{\IEEEauthorblockN{Daniel Lersch\IEEEauthorrefmark{1},
Malachi Schram\IEEEauthorrefmark{1},
Zhenyu Dai\IEEEauthorrefmark{1}, 
Kishansingh Rajput\IEEEauthorrefmark{1},
Xingfu Wu\IEEEauthorrefmark{2},
N. Sato\IEEEauthorrefmark{3},
J. Taylor Childers\IEEEauthorrefmark{4}
}
\IEEEauthorblockA{\IEEEauthorrefmark{1}Data Science Department,
Thomas Jefferson National Accelerator Facility, Newport News, VA 23606 USA}
\IEEEauthorblockA{\IEEEauthorrefmark{2}Mathematics and Computer Science Division, Argonne National Laboratory, Lemont, IL 60439 USA}
\IEEEauthorblockA{\IEEEauthorrefmark{3} Theory Center,
Thomas Jefferson National Accelerator Facility, Newport News, VA 23606 USA}
\IEEEauthorblockA{\IEEEauthorrefmark{4}Leadership Computing Division, Argonne National Laboratory, Lemont, IL 60439 USA}
}

\begin{document}

\maketitle

\begin{abstract}
Large scale, inverse problem solving deep learning algorithms have become an essential part of modern research and industrial applications. 
The complexity of the underlying inverse problem often poses challenges to the algorithm and requires the proper utilization of high-performance computing systems. 
Most deep learning algorithms require, due to their design, custom parallelization techniques in order to be resource efficient while showing a reasonable convergence. 
In this paper we introduces a \underline{S}calable \underline{A}synchronous \underline{G}enerative workflow for solving \underline{I}nverse \underline{P}roblems \underline{S}olver (SAGIPS) on high-performance computing systems. We present a workflow that utilizes a parallelization approach where the gradients of the generator network are updated in an asynchronous ring-all-reduce fashion. Experiments with a scientific proxy application demonstrate that SAGIPS shows near linear weak scaling, together with a convergence quality that is comparable to traditional methods. The approach presented here allows leveraging GANs across multiple GPUs, promising advancements in solving complex inverse problems at scale.
\end{abstract}

\section{Introduction}
\label{sec_intro}
This work presents a generative, inverse problem solving workflow which runs across multiple GPUs of a high performance computing system. 
The workflow was initially developed for the SciDAC QuantOm project~\cite{quantom22}, however, it's designed to provide a generic optimization and control solution which include inverse problems. 

The two main characteristics of the SciDAC QuantOm project that require a distributed workflow are:
\begin{enumerate}
    \item Large volume data files, each containing $~\sim 10^{10}$ of physics events. 
    \item The workflow components themselves (see Section~\ref{sec_framework}) require an $\mathcal{O}(\text{exaFLOPS})$ computational resources in order to analyze the input data produced at the Electron Ion Collider~\cite{accardi2014electron}. 
\end{enumerate}
It is expected, that the main contribution to item 2 is the stochastic event sampler which ensures that the predicted data matches the format (i.e. events) of the input data. The stochastic nature of the event sampler poses a first challenge for running the workflow in a distributed manner.

The second challenge results from the Generative Adverserial Neural Network (GAN) which defines the optimization part of the workflow (see Section~\ref{gan_intro}). GANs consist of two neural networks that need to be trained in a synchronized manner, in order to generate meaningful data. Parallelizing the training of GANs presents significant challenges due to their non-stationary objective function~\cite{salimans2016improved}, frequent synchronization requirements between the generator and discriminator networks~\cite{10121444}, and the risk of mode collapse~\cite{salimans2016improved,hoang2018mgan}. These complexities are further amplified when scaling GAN training across multiple devices or nodes.

There are already tools available, such as FeGAN~\cite{fegan2020} or MDGAN~\cite{mdgan2019}, which allow to efficiently train a GAN accross multiple GPUs. Unfortunately, these are not viable options for our workflow, because the GAN is not utilized in the traditional sense. 
Unlike in a regular GAN, the generator predictions are not directly fed into the discriminator, rather passed through a differentiable, modular, pipeline. 
The outcome of this pipeline is then prompted to the discriminator. 
An essential part of that pipeline is the sampling module mentioned above. 
The underlying inverse problem, that one tries to solve, dictates the complexity of the pipeline and therefore has an impact on the parallelizability of the GAN training itself. 
As we highlight the details of our workflow, we will elaborate on the need of a custom distributed training strategy. 
Additionally, we will  show that existing tools, such as horovod~\cite{hvd2018}, are not applicable to our use-case. 

We present a method that is similar to~\cite{mdgan2019}, however,  we use multiple distinct discriminator networks that learn autonomously. 
Instead, we send the initial copies of the generator weights to each rank and update the generator gradients in an asynchronous ring-all-reduce fashion. 
Meaning that no rank is preferred over the other, as apposed to a master-worker system. 
Our approach may be seen as some sort of hybrid between~\cite{fegan2020} and~\cite{mdgan2019}. 
We furthermore explore the means of transferring the gradients between the individual GPUs, by utilizing remote memory access. 
We will also show that we are able to reduce the communication overhead a bit further, by introducing a grouping mechanism.

\section{The SAGIPS Workflow}
\label{sec_framework}
The main purpose of this workflow is to solve inverse problems. However, it is designed such that it can also solve generic optimization and control problem as well. The following is an example inverse problem that can be solved with our approach: Suppose an observable ${\bm y}$ that is obtained through measurements (e.g., temperature). The object of interest however is a set of features ${\bm x}$ which describes the underlying nature of ${\bm y}$ (e.g., Brownian motion of molecules). Unfortunately, the features ${\bm x}$ are neither directly accessible nor measurable, but the relation between ${\bm x}$ and ${\bm y}$ is known via a function $f$:
\begin{equation}
    \bm {y} = f(\bm{x})
    \label{def_inv}
\end{equation}
The complexity of $f$, however, does not allow it to be inverted trivially. 
This problem is overcome by approximating ${\bm x}$ itself\footnote{With an adequate theoretical model ${\bm \hat{x}}$, that is.}, such that for a given set of parameters ${\bm p}$ we obtain:
\begin{equation}
    {\bm x} \approx {\hat{\bm x}}({\bm p})
    \label{def_x_hat}
\end{equation}
Now we are able to solve \eqref{def_inv} by minimizing an objective Function $F$ (such as log-likelihood, etc.) with respect to ${\bm p}$:
\begin{equation}
    \min_{p} F[{\bm y},f({\hat{\bm x}}({\bm p}))]
    \label{def_optim}
\end{equation}
Once the parameters ${\bm p}$ are found, we can extract ${\bm x}$ via \eqref{def_x_hat}. 
The SAGIPS workflow is basically a computational manifestation of \eqref{def_optim}. 

\begin{figure}[htbp]
\centering
\includegraphics[width=0.47\textwidth]{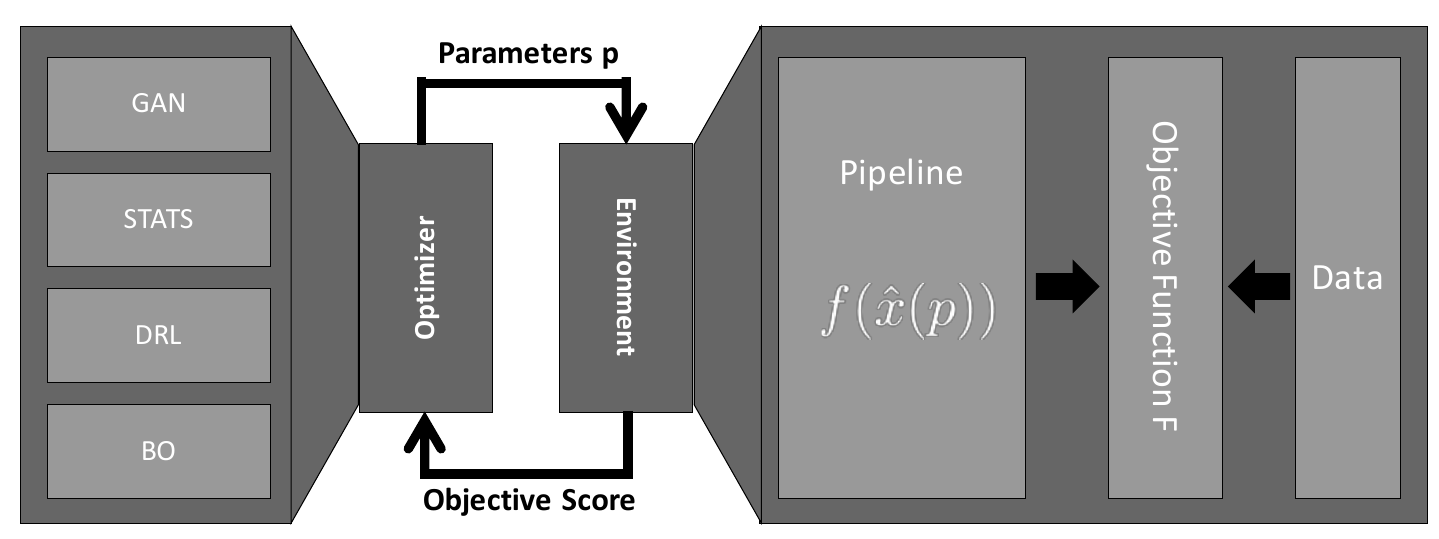}
\caption{Schematic representation of the SAGIPS workflow with all its modules and dependencies. The individual models are described in the text.}
\label{fig:wrkfl}
\end{figure}

In the following, we will briefly describe the core features of our workflow, as Fig.~\ref{fig:wrkfl} provides a schematic overview of SAGIPS. 
The modus operandi is as follows: The optimizer predicts a set of parameters ${\bm p}$ that are passed to an environment. 
The environment, directly corresponds to $F[{\bm y},f({\hat{\bm x}}({\bm p}))]$ in \eqref{def_optim}, internally translates the parameters to an objective score (e.g., log-likelihood) which is sent back to the optimizer. 
The optimizer utilizes the objective score to update its internal state (i.e. the optimizer is trained) and responds with a new set of parameters to the environment. 
This interaction continues until the objective score is minimized (or maximized, depending on the objective function). 

\subsection{The Optimizer}
The optimizer is any trainable algorithm that predicts new parameters based on the incoming objective score. 
Techniques that can be used within the workflow includes,  but not limited to, GANs, Bayesian Optimization (BO), Deep Reinforcement Learning (DRL), and other statistical methods. 
Examples of such algorithms are given in the far left box of Fig.~\ref{fig:wrkfl}. 
In this work a GAN optimizer is used, which will be explained in the following section.

\subsection{The Environment}
The design of the environment is ultimately defined by the underlying inverse problem that one tries to solve. 
In the context of this paper, the environment is set up to analyze a toy problem which will be explained later in more detail in this paper.
The environment is composed of two key elements:
\subsubsection{Pipeline}
This part of the environment corresponds to $f({\hat{\bm x}}({\bm p}))$ in \eqref{def_optim}. The pipeline translates the predicted parameters ${\bm p}$ into a synthetic data set $f(\hat{\bm x}({\bm p}))$ which is compatible with the reference data ${\bm y}$. The pipeline itself consists of multiple, independent modules of which one is a sampler. The sampler ensures that the synthetic data matches the format of the reference data (e.g., individual events or sequences). 

\subsubsection{Objective Function} 
This module represents $F$ in \eqref{def_optim} and compares the synthetic data to the input (or original) data by returning a score. 
One important constraint on the objective function is that it has to be able to compare mismatched data sets (mismatched w.r.t to the entry index). 
The reference data can not be directly compared to the synthetic data, due to the random order of the generated sampled data.
A mean squared error objective for example will not be helpful, as it relies on a direct comparison between two data sets (i.e. entry i in the reference data is compared to entry i in the synthetic data). In the case of a GAN optimizer, the objective function is a discriminator neural network which is trained to label the reference data as one and the synthetic data as zero. The discriminator does therefore not require a matching between the two data sets.

\subsection{Distributed Analysis}
\label{sec_dist_ana_constriants}
As laid out in~\ref{sec_intro}, distributed training is imperative for our workflow. Another aspect, that we will not address in this paper\footnote{Simply because the SciDAC QuantOm project is in its early stages.}, is that the SAGIPS workflow needs to analyze multiple reference data sets in parallel, or simultaneously fit multiple observables within one data set. In either case, the environment pipeline is different\footnote{For different reference data sets, or observables, we may have to use different pipeline settings.} for each data set / observable, but the predicted parameters remain the same. 

\subsection{Software Environment and Hardware}
We conduct our experiments on Polaris~\cite{alcfpolaris} at Argonne National Laboratory. The core packages are Horovod 0.28.1, mpi4y 3.1.5, PyTorch 2.0.1, Python 3.10 and cudatoolkit-standalone 11.8.0. 

\section{GAN Optimizer}
\label{gan_intro}

\subsection{GAN in a Nutshell}
\label{gan_nutshell}
\begin{figure}[htbp]
    \centering
    \includegraphics[width=0.5\textwidth]{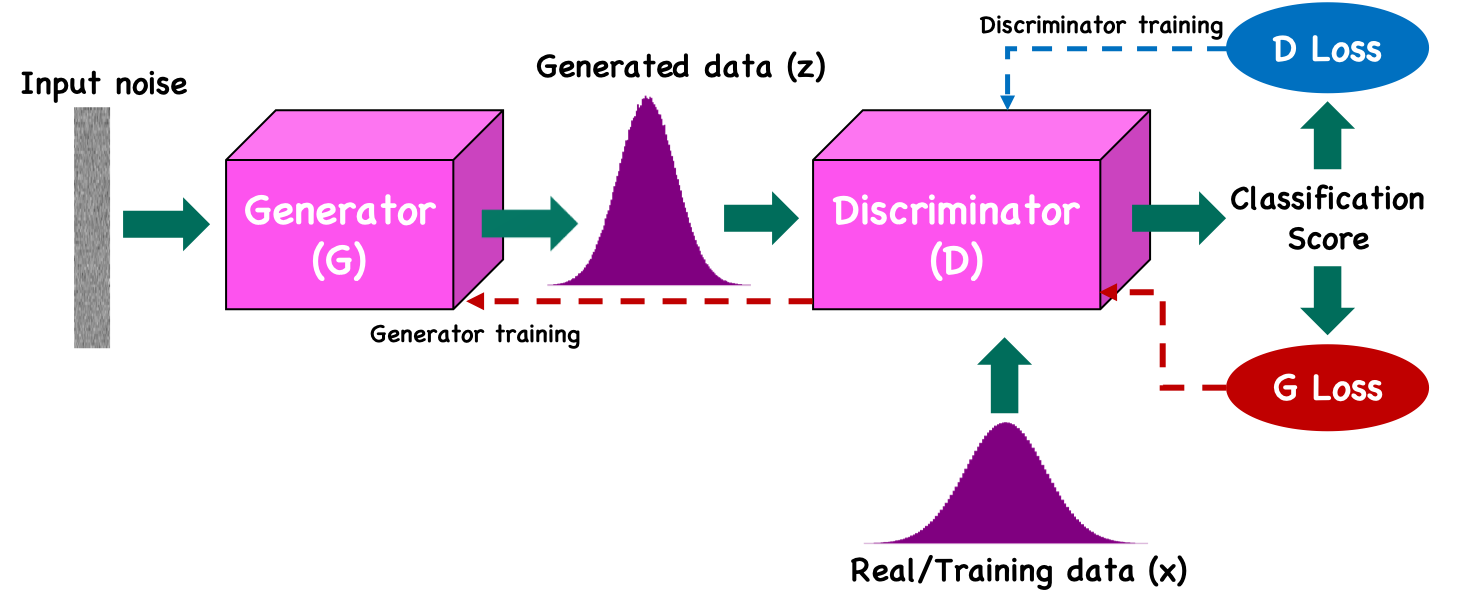}
    \caption{Schematic representation of training a GAN.}
    \label{fig:gan_training}
\end{figure}

GANs~\cite{goodfellow2014generative,mirza2014conditional} are a class of artificial intelligence algorithms used in unsupervised machine learning while using a supervised loss. The fundamental idea behind a GAN is to train two neural networks, namely the generator and the discriminator, in a competitive setting (see Fig.~\ref{fig:gan_training}). 

\textbf{Generator}:  This network learns to produce synthetic data, like images or text, from random noise. It aims to generate samples indistinguishable from real data. During training, it adjusts its parameters to minimize the discriminator's ability to distinguish between real and fake data.

\textbf{Discriminator}: The discriminator distinguishes between real and fake data generated by the generator. It acts as a binary classifier, assigning probabilities to samples being real or fake. Trained on a mix of real and fake data, it improves at discerning real from fake samples over time.

Both the generator and the discriminator rely on each other for training. The generator seeks to produce realistic samples to deceive the discriminator, while the discriminator aims to become better at distinguishing between real and fake samples. This adversarial process drives both networks to improve, resulting in the generation of high-quality synthetic data mimicking real data distributions.

\subsection{Deep Learning Framework}
So far, all SAGIPS modules are written in PyTorch~\cite{torch2019} and we are using the built-in automatic differentiation engine (autograd) to keep track of the gradient propagation through each individual module. PyTorch requires that all tensors are explicitly loaded into GPU or CPU memory. This can be helpful, as it allows controlling the GPU resources at each stage of the workflow training process.

\subsection{Parallelization Strategy}
\label{sec_strategy}
Early implementations of our workflow tried to train both the generator and discriminator, across multiple GPUs, similar to~\cite{fegan2020}. The observed scaling behavior was not promising so we ended further investigation in that direction. Similar conclusions were drawn when we considered the idea of a centralized generator with multiple discriminators, as done in~\cite{mdgan2019}. The following section will highlight the details of our parallel training strategy. 

\section{Distributed Training of the GAN Workflow}
\label{sec_distributed}
Within this work, we examined two options for training the GAN workflow on a HPC system: (i) Ensemble Analysis and (ii) Asynchronous Data Parallel Training. The main difference between option (i) and (ii) is the communication between the individual GAN workflows. While the former uses no communication (i.e. the GANs are trained independent of each other on a single GPU), the latter transfers the generator gradients between workflows, where each workflow runs on its own GPU. From now on, we will use the terms GPU and rank interchangeably.

\subsection{Ensemble Analysis}
Ensemble methods have emerged as a prevalent strategy for augmenting the efficacy of GANs~\cite{lakshminarayanan2017simple, tolstikhin2017adagan, han2020gan}. This section is dedicated to scrutinizing the effectiveness of standard ensembles of GANs in addressing bias and variance. We evaluate models configured with different settings. Our objectives are to illustrate: (i) That larger models featuring increased model parameters within the GAN architecture, coupled with augmented training data, exhibit enhanced performance overall. (ii) We ascertain the potential of scalability for GANs through ensemble sizes up to 100. These findings elucidate the necessity for scaling GANs, substantiating the imperative of extending their performance beyond what individual models can achieve.

\subsection{Asynchronous Data Parallel Training with Overlap}
The main idea behind data parallel training is that each GPU receives a copy of the model to be trained and only a portion of the total available data set is processed. 
\begin{figure}[htbp]
    \centering
    \includegraphics[width=0.45\textwidth]{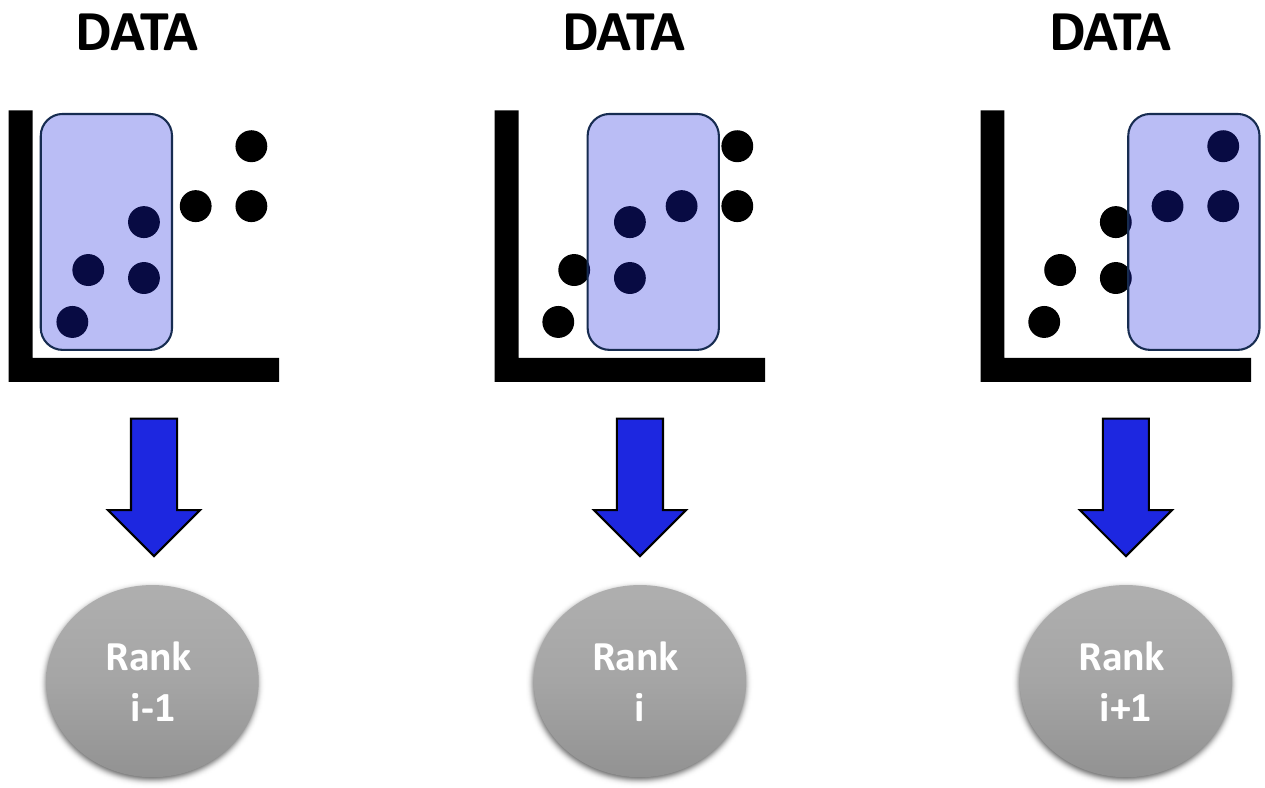}
    \caption{Distributing a common data set across multiple ranks. Each rank has its own copy of the data, but analyzes a fraction only (indicated by transparent rectangle).}
    \label{fig:dat_shard}
\end{figure}
This work uses, at its core, the same method with a few alterations. Each GPU has a copy of the generator network, but trains its own discriminator locally. The data used to train the GAN is loaded by the master rank (rank 0) and distributed to all other ranks\footnote{This will change in future applications, due to the expected data size mentioned earlier.}. Each rank then analyzes a (randomly drawn) sub-fraction of its data (see Fig.~\ref{fig:dat_shard}). Ideally, this leads to a faster convergence, because the GAN is effectively processing more data at once.

Every rank randomly draws training sub-samples (via bootstrapping) from its data and feeds them through the GAN. The discriminator gradients are updated right away whereas the generator gradients are transferred to neighbouring ranks.

\subsubsection{Gradient Transfer}
The crucial element of data parallel training is the gradient transfer between the model copies at each rank. Without it, the method above would be another ensemble analysis. The gradient transfer ensures that the models at each rank exchange information regarding the next optimization step, which ideally improves the convergence rate. 

\subsubsection{Asynchronous Ring-All-Reduce}
\label{sec_ring_allred}
There are different ways to communicate gradients between multiple GPUs, such as the Hierarchical-Reduce~\cite{group2018}, the 2D-Torus-all-reduce~\cite{torusring2019} or double binary trees~\cite{binarytree2013}. The latter has been proven to be superior to all ring-based communication methods~\cite{binarytree2013}. This work, however, focuses on the asynchronous ring-all-reduce technique because it is a well-established and straightforward method to implement. We also wanted to ensure initially that our workflow is parallelizable before exploring different communication techniques.

\begin{figure}[htbp]
\centering
 \includegraphics[width=0.4\textwidth,origin=l]{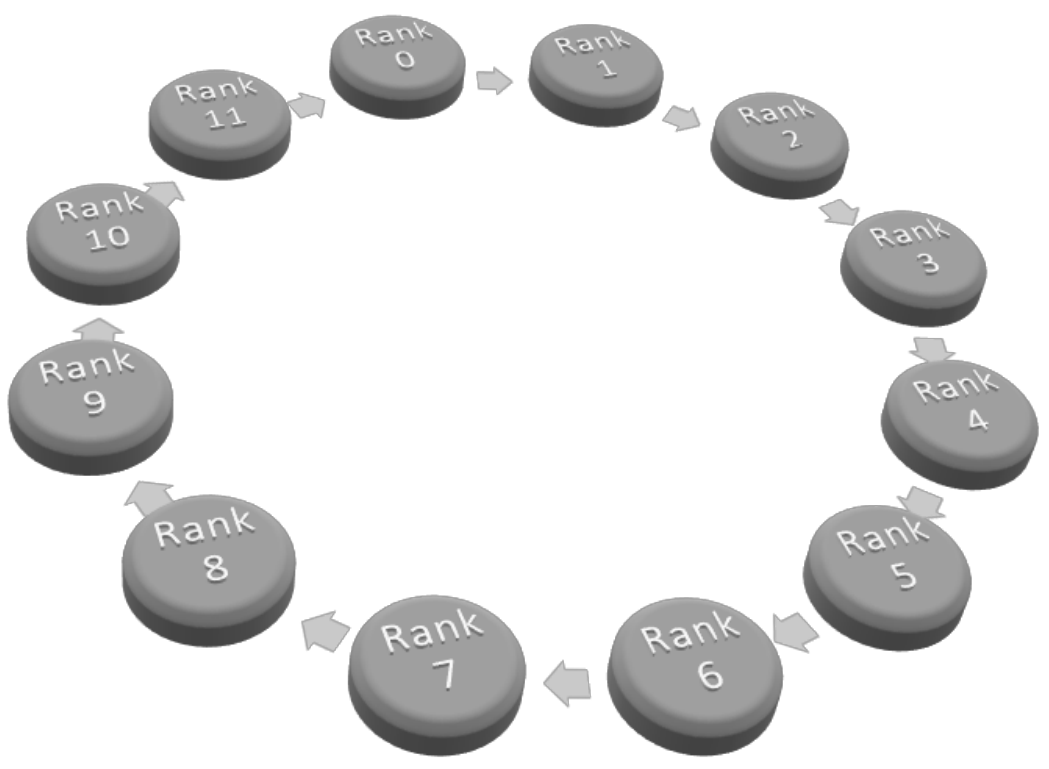} 
 \caption{Schematic representation of a ring-all-reduce communication between 12 ranks.}
 \label{fig:ring_all_red}
\end{figure}

Fig.~\ref{fig:ring_all_red} displays such a ring-all-reduce communication between twelve ranks.

The basic idea behind this procedure is that each rank communicates with its adjacent neighbour only (see pseudo code~\ref{algo_rar}).
\begin{algorithm}
  \caption{Ring-All-Reduce Communication}
  \label{algo_rar}
  \begin{algorithmic}
   \REQUIRE $N$ ranks (GPUs), each having access to the gradient tensors ${\bm g}$ of their generator network.
   \STATE
   \FOR{i=1,...,$N$-1}
      \STATE Rank i sends gradients ${\bm g}_i$ to Rank i+1
      \STATE Rank i receives gradients ${\bm g}_{i-1}$ from Rank i-1
      \STATE Update gradients on Rank i: ${\bm g}_i \mapsto {\bm g}_i + {\bm g}_{i-1}$
   \ENDFOR
  \end{algorithmic}
 \end{algorithm}
The advantage of this method is that Rank i only sends data to Rank i+1 or receives data from Rank i-1. It does not need to communicate with all remaining ranks in the system. This reduces the communication overhead, compared to a system where all the information is accumulated and distributed back via a single (master) node. The gradient transfer between two adjacent ranks is done asynchronously by using the mpi4py library.

The current implementation does not divide the gradient tensors into chunks, in order to further optimize the communication. This will be part of future investigations.

\subsubsection{Remote Memory Access - RMA}
\label{sec_rma}
The crux of the SAGIP workflow is that it relies on a pipeline which translates the generator predictions into meaningful data. Depending on the pipeline, the sampling process can be very time intensive (We observed up to $1\,\rm{min}$ per epoch for one specific pipeline prototype). A communication between multiple ranks can suffer from this, as some ranks may run the data generation task faster / slower than others. For a ring-all-reduce communication, this means Rank i has to wait for Rank i+1 to finish first, before it is open for communication. 
\begin{figure}[htbp]
\centering
 \includegraphics[width=0.45\textwidth,origin=r]{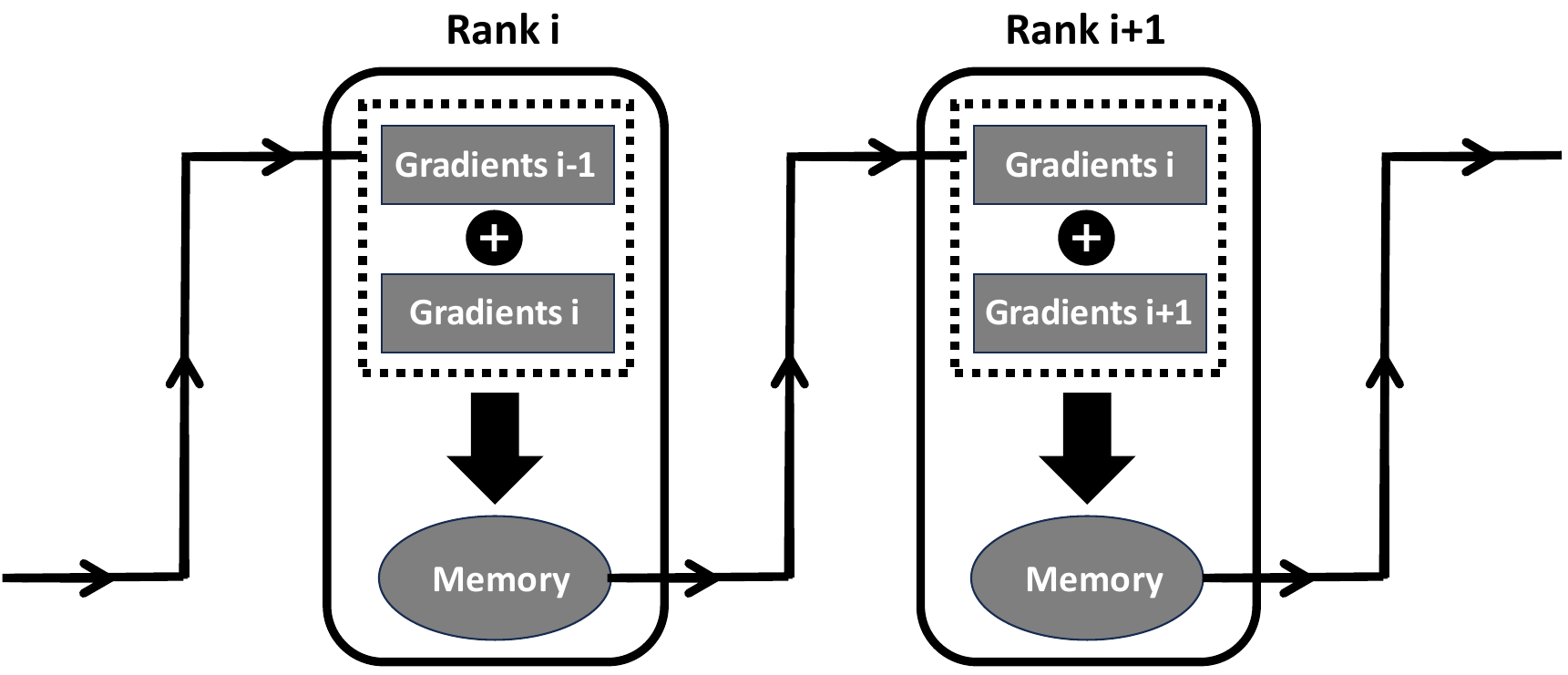}
 \caption{Schematic representation of a Remote Memory Access communication between two ranks $i$ and $i+1$.}
 \label{fig:rma_scheme}
\end{figure}
For this reason, we studied the feasibility of using Remote Memory Access (RMA) communication instead (see Fig.~\ref{fig:rma_scheme}). RMA enables a rank to either write gradients to or read gradients from the memory of an other rank. This has the advantage that a given rank does not have to wait for an other rank to finish its current task before gradients can be sent / received.  The gradients to be transferred are just written to memory and then fetched by the other rank whenever it is ready. 

For the purpose of this work, the ranks still communicate in a ring-all-reduce fashion with each other, but instead of sending / receiving gradients to / from neighbouring ranks, they will be written to / loaded from memory.

\subsubsection{Grouping}
\label{sec_grouping}
\begin{figure}[htbp]
    \centering
    \includegraphics[width=0.45\textwidth]{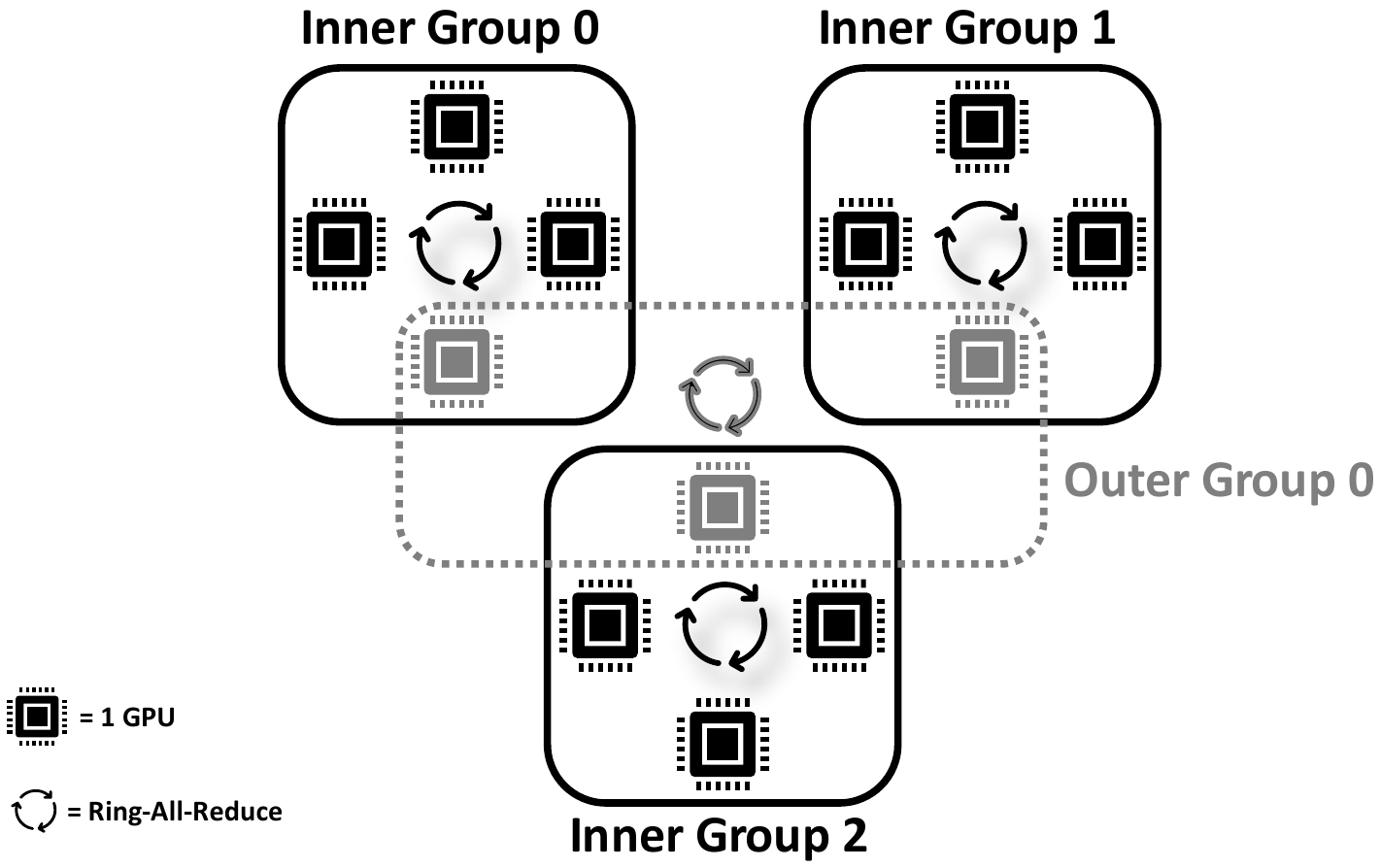}
    \caption{Grouping 12 ranks into three inner groups and one outer group.}
    \label{fig:group_scheme}
\end{figure}

In order to further optimize the communication between multiple ranks, we introduced a grouping scheme (see Fig.~\ref{fig:group_scheme}). Instead of applying the ring-all-reduce to multiple GPUs across different nodes, we divide the available ranks into groups (called inner groups from now on) and restrict the ring-all-reduce communication to the members of that group. The idea of grouping is not new and has already been successfully utilized in~\cite{group2018}. 

The key difference between our implementation and the Hierarchical All-Reduce in~\cite{group2018} is that we do not use a three step communication and do not rely on broadcasting gradients from a master rank. The details of our grouping mechanism are explained below.
The size of an inner group is (for now) defined by the number of available GPUs on a computing node. For example: If we have three computing nodes with four GPUs each, then this would lead to three individual inner groups with four ranks (see solid, black rectangles in Fig.~\ref{fig:group_scheme}). Each inner group uses its own ring-all-reduce for communication, or in other words: Only ranks that share the same physical node transfer and accumulate gradients amongst each other.  

We will show in the following sections that this method indeed reduces the communication overhead and thus allows to run an analysis with more ranks within a given time frame. However, the setting described so far has one drawback: The gradients are transferred between ranks on one node only, but not between nodes. We solve this issue by introducing an outer group which takes in one rank from each inner group (see gray, dashed rectangle in Fig.~\ref{fig:group_scheme}). In case of our example with 12 GPUs, this means we have one outer group with three GPUs. The members of this outer group transfer gradients amongst each other. This ensures that gradients are also shared across nodes.

The inner group communication happens every epoch, whereas the outer group communication happens at a specified frequency, see Tab.~\ref{tab:group_summary}. This allows controlling the communication between nodes.  
\begin{table}[htbp]
    \centering
    \caption{Settings for grouping the communication between multiple ranks across different nodes. The outer group update frequency $h$ is a hyperparameter that can be adjusted by the user.}
    \begin{tabular}{c||c|c}
            & Size & Update Frequency \\
    \hline
    \hline
    Inner Group  &  $\#$ GPUs on Node & Every Epoch\\
    Outer Group  & $\#$ inner Groups & Every h epochs 
    \end{tabular}
    \label{tab:group_summary}
\end{table}
The rank chosen from each inner group for the outer group communication is fixed to be rank 0. We envision this choice to be a random process in future implementations. 

\subsubsection{Available Modes}
\label{sec_modes}
The methods described above allow us to train the GAN in different ways, or modes, as shown in Tab.~\ref{tab:grad_modes}. 
\begin{table}[htbp]
    \centering
    \caption{Available modes for training the GAN across multiple ranks. The abbreviation {\bf ARAR} refers to the {\bf A}synchronous {\bf R}ing-{\bf A}ll-{\bf R}educe explained in section~\ref{sec_ring_allred}. {\bf RMA-ARAR} on the other hand refers to {\bf A}synchronous {\bf R}ing-{\bf A}ll-{\bf R}educe, but the communication is done via {\bf R}emote {\bf M}emory {\bf A}ccess.}
    \begin{tabular}{c||c|c|c|}
    Mode & Inner Group & Outer Group & No Group \\
         & Communication & Communication & \\ 
    \hline
    \hline
    ARAR & none & none & ARAR \\
    RMA-ARAR-ARAR & RMA-ARAR & ARAR & \\
    ARAR-ARAR & ARAR & ARAR & \\
    \end{tabular}
    \label{tab:grad_modes}
\end{table}
The first mode shown in Tab.~\ref{tab:grad_modes} uses "conventional" asynchronous ring-all-reduce without grouping. This means that all ranks talk to each other in a ring-like fashion. The second and third mode use grouping as described in section~\ref{sec_grouping}. The only difference is that the second mode use the RMA based ring-all-reduce for the inner group communication. 

\subsubsection{Gradient Off-Loading}
All modes described in Tab.~\ref{tab:grad_modes} run on CPU, i.e. the generator gradients on each rank are loaded off the GPU and into the CPU memory. Then, after accumulating gradients via one of the modes in Tab.~\ref{tab:grad_modes}, the gradients are registered back into the GPU memory, so that the generator weights can be updated. This gradient off- and on-loading helps to regulate the GPU memory footprint, especially because we are using PyTorch and the gradient tensors of each SAGIPS module\footnote{Each SAGIPS module needs to be differentiable, otherwise we would not be able to train a GAN via backpropagation.} are tracked by PyTorch's autograd. 

\subsection{Software Libraries for Transferring Gradients}
\label{sec_soft_lib}
All gradient transfer modes listed in Tab.~\ref{tab:grad_modes} are managed by the MPI4Py library~\cite{mpi_4py}. The reason for this is twofold: (i) It supports Python and Python Numpy data and (ii) It provides all the tools we need to enable the communication between ranks. 

It should be noted that PyTorch has its own library for distributed data parallel training~\cite{torchdist2020}. Even though the current version of SAGIPS is written in PyTorch, future implementations may prefer different deep learning frameworks such as TensorFlow~\cite{tf2015}. In order to maintain the generality of the workflow, we decided to favor mpi4py which is agnostic to any deep learning framework. Plus, it provides the flexibility we need to customize any parallelization method to our workflow. 

\section{Experiments}
\label{sec_experiments}
\begin{figure}[htbp]
    \centering
    \includegraphics[width=0.5\textwidth]{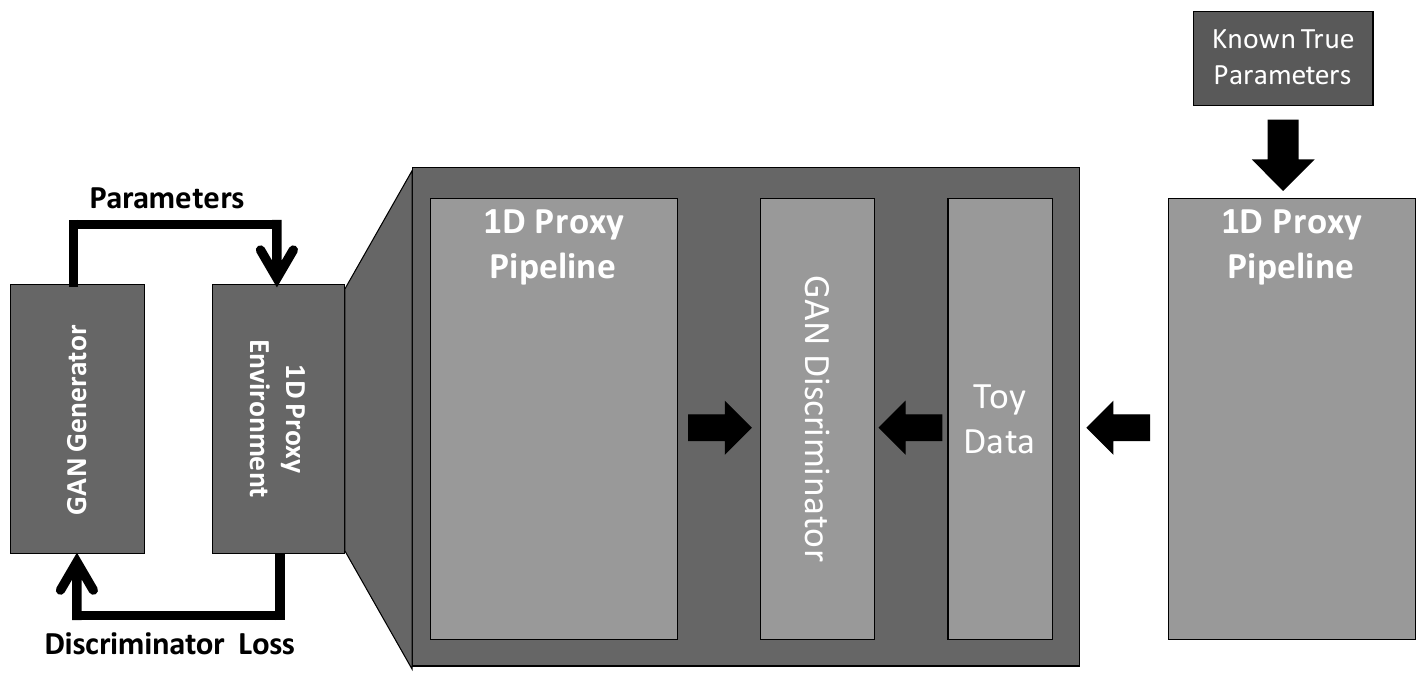}
    \caption{Schematic representation of the loop closure test that was used to run the scaling experiments. Please see text for details.}
    \label{fig:loop_close}
\end{figure}

The scaling experiments conducted for this work are based on a loop-closure test depicted in Fig.~\ref{fig:loop_close}. The SAGIPS optimizer is a GAN generator which means that the objective function is a GAN discriminator. The pipeline $f(\hat{\bm x}({\bm p}))$ within the environment (referred to as proxy 1D app in the introduction) translates six parameters $p_0,...,p_5$ to two observables:
\begin{equation}
    \label{def_proxy}
    f(\hat{\bm x}(p_0,...,p_5)) = (y_0,y_1)
\end{equation}
The exact same pipeline is used, together with a set of 6 known parameters to create a toy data set, consisting of $(y_0,y_1)$. The SAGIPS workflow uses the GAN generator to predict a set of parameters $\hat{p}_0,...,\hat{p}_5$ which are fed through the pipeline and lead to:
\begin{equation}
    \label{def_proxy_pred}
    f(\hat{\bm x}(\hat{p}_0,...,\hat{p}_5)) = (\hat{y}_0,\hat{y}_1)
\end{equation}
The discriminator is trained to distinguish between $(y_0,y_1)$ and $(\hat{y}_0,\hat{y}_1)$. The generator is trained to produce a set of parameters $\hat{p}_0,...,\hat{p}_5$ that minimizes the difference between the discriminator response on \eqref{def_proxy} and \eqref{def_proxy_pred}. Since both, the toy data set and the SAGIPS environment, rely on the same 1D proxy app pipeline, the predicted and input parameters are expected to be equal: $p_i-\hat{p}_i=0, \forall i=0,..,5$. We therefore use the agreement between predicted and known parameters as a measure of convergence quality for the GAN.

\subsection{Running the GAN Workflow}
\label{sec_gan_workflow_settings}
Tab.~\ref{tab:gen_settings} summarizes the settings that were used to run the GAN workflow for either the ensemble or the asynchronous data parallel training. Both the generator and discriminator are trained for $100\,\rm{k}$ epochs. The generator predicts $1024$ parameter samples (see third row in Tab.~\ref{tab:gen_settings}). These parameters are passed through the 1D proxy app pipeline which samples 100 events per parameter prediction (see fourth row in Tab.~\ref{tab:gen_settings}). This leads to a total of $102,400$ synthetic events. This number ultimately defines the batch size for the discriminator, because the number of synthetic and real data samples needs to be identical. 
\begin{table}[htbp]
    \centering
    \caption{General settings for running the GAN workflow.}
    \begin{tabular}{c||c}
    \hline
       Number of epochs  &  $100\,\rm{k}$\\
       Discriminator Batch Size  & $102,400$ \\
       $\#$ Predicted Parameter Samples & $1024$ \\
       $\#$ Events generated per Parameter Sample & $100$\\
  \hline
    \end{tabular}
    \label{tab:gen_settings}
\end{table}
The sampler used within the 1D proxy app relies on the inverse CDF method, i.e. we use the inverse of a differentiable function to sample events from a given one dimensional distribution. The choice of this sampler was based on (a) differentialbility and (b) simplicity. The latter was important for efficient prototyping and testing SAGIPS. The former is crucial for using backpropagation for the GAN training.

In the ensemble study, we explored variations in the number of model parameters and training batch sizes. However, in the subsequent scaling section, all hyperparameters were kept constant. The generator has a total of $51,206$ trainable parameters, whereas the discriminator has $50,049$ parameters. Both networks use Leaky ReLU activation functions in the hidden layers, together with a Kaiming normal weight initialization. The generator learning rate was set to $10^{-5}$, whereas the discriminator learning was $10^{-4}$ respectively. These GAN settings were found by manual tuning. 

\subsection{Experiments with Ensemble Training}
We tested GANs with different number of model parameters and training batch size. Each run of every model is independently trained for 100 k epochs on a single GPU. Subsequently, each model is trained 20 times to form an ensemble. Additionally, the model with the largest model parameters and batch size undergoes further training for an additional 80 times to form an ensemble size of 100, facilitating the evaluation of ensemble methods at a larger scale.

\subsection{Experiments with Asynchronous Data Parallel Training}
We test the distributed training methods described in section~\ref{sec_distributed}, by running the GAN workflow in the three modes which are summarized in Tab.~\ref{tab:grad_modes}. For all experiments, we set the update frequency $h$ for the communication between nodes to: $h=1\,\rm{k}$. Which means that the gradients are shared across nodes every $1,000$-th training epoch. We found this setting by running an analysis with 200 GPUs and different settings for $h$. The value reported here corresponds to the best parameter convergence (i.e. $p_i-\hat{p}_i$) as a function of time. 

Furthermore, we restrict the gradient transfer to the generator weights, i.e. only the gradients of the generator weights are shared.
The bias gradients are one dimensional tensors and known to slow down the ring-all-reduce communication~\cite{group2018}. In our analysis, we also noticed that adding the bias gradients does not noticeably improve the GAN training. We do plan however to investigate the feasibility of using tensor fusion which allows to combine small tensor into a larger one.

All experiments were conducted on the Polaris machine~\cite{alcfpolaris} at Argonne National Lab. Each Polaris node uses one AMD EPYC "Milan” processor and 4 NVIDIA A100 GPUs, together with a HPE Slingshot 11 network.

\section{Results}
\label{sec_results}
We evaluate the experiments presented earlier, with respect to two criteria: (i) Overall Training Time and (ii) Convergence quality. We define the latter by comparing the input parameters $p_{i}$ to the parameters predicted by the generator model $\hat{p}_{i}$ via normalized residuals:
\begin{equation}
     \label{def_param_res}
    \hat{r}_i = \frac{p_{i} - \hat{p}_{i}}{p_{i}}
\end{equation}
This quantity as a function of training time has proven to be a better indicator for convergence than the corresponding GAN loss curves. Various tests showed that the GAN loss curves indicate convergence, but the actual parameter residuals have not converged towards zero yet. This is a feature of our workflow, because, depending on the pipeline, a non-optimal parameter prediction may still lead to a reasonable agreement between the input and synthetic observables. 

\subsection{Computing the Ensemble Response}
\label{sec_ensemble_res}
Given a single noise vector ${\bm n}$ and $M$ trained generator networks $G_i$ $i=1,...,M$, we define the ensemble parameter prediction as:
\begin{equation}
    \hat{\bm p} = \frac{1}{M}\sum\limits_{i=1}^{M}G_i({\bm n})
    \label{def_ens_mean}
\end{equation}
Consequently, we deduce the uncertainty from each parameter prediction:
\begin{equation}
    {\bm \sigma} = \sqrt{\frac{1}{M} \sum\limits_{i=1}^{M}\Big [G_i({\bm n})-\hat{\bm p}\Big ]^2}
    \label{def_ens_std}
\end{equation}
For a batch of $k$ noise vectors we simply report the average of $\hat{\bm p}$ and ${\bm \sigma}$ across the batch dimension $k$. If the GAN networks within the ensemble were trained via one of the methods listed in Tab.~\ref{tab:grad_modes}, we simply average over the predictions from all generators and then insert that average into \eqref{def_ens_mean} and \eqref{def_ens_std}.

\subsection{Results from the Ensembles Analysis}
\begin{figure}[htbp]
    \centering
    \includegraphics[width=0.48\textwidth]{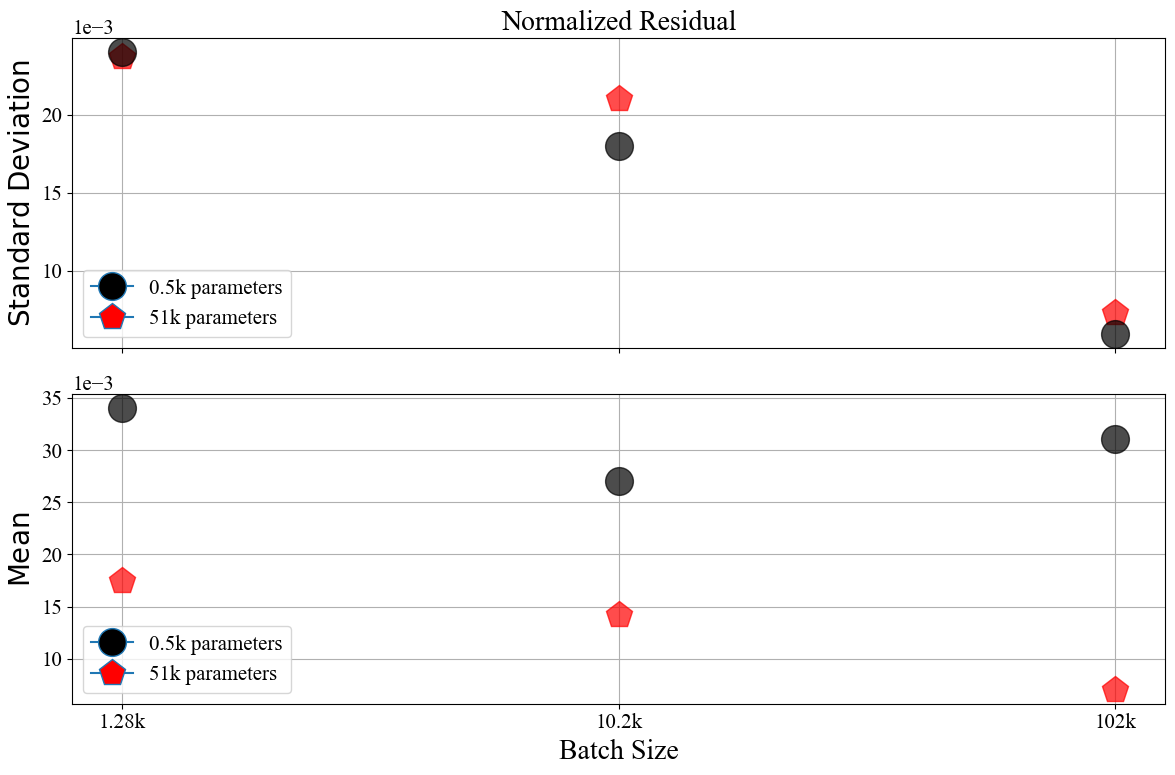}
    \caption{Standard Deviation ({\bf top}) and mean ({\bf bottom}) of the normalized residual $\hat{r}_0$. These results are derived from each model being trained 20 times to generate an ensemble prediction based on (\ref{def_param_res})(\ref{def_ens_mean})(\ref{def_ens_std}).}
    \label{fig:ensemble_analysis}
\end{figure}

Fig.~\ref{fig:ensemble_analysis} presents a summary of ensemble results derived from 20 GANs trained with varying numbers of parameters and datasets. The bottom panel illustrates that larger models trained with more extensive datasets exhibit smaller normalized residuals at the conclusion of training. This outcome aligns with expectations, as models with increased parameter counts possess greater capacity to mitigate bias and generalize effectively across diverse training data. In addition, the top panel reveals that models trained with limited datasets encounter data scarcity issues, resulting in higher uncertainties. 

\begin{figure}[htbp]
    \centering
    \includegraphics[width=0.48\textwidth]{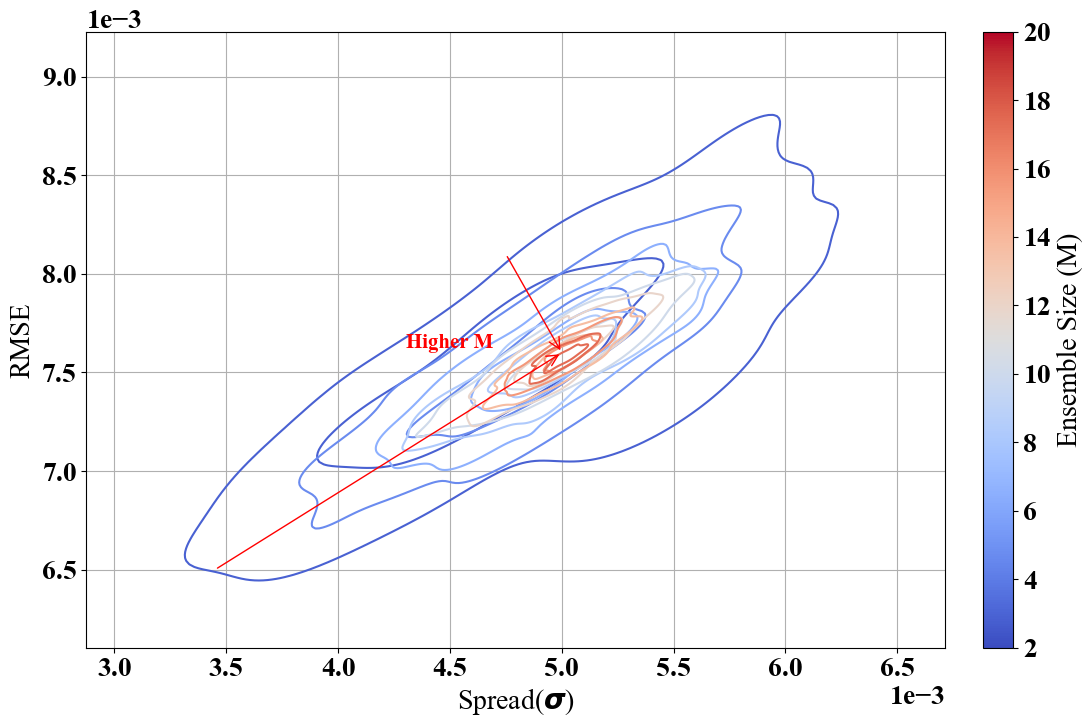}
    \caption{The 95\% confidence level contour plot of root mean square error (RMSE) versus spread ($\sigma$)for ensemble sizes ranging from 2 to 20. Red arrows indicate the direction of increasing ensemble size (M). Data are derived from $M$ runs sampled from a total of 20 ensemble runs of GAN with 51,200 parameters, utilizing a batch size of 102,000.}
    \label{fig:rmse_spread_model1}
\end{figure}

To further investigate the relationship between residual and standard deviation across different ensemble sizes, we sampled various ensemble sizes ($M$) from the ensemble of 20 runs of model with 51k parameters and batch size 102k, as shown in Fig.~\ref{fig:rmse_spread_model1}. We conducted 300 samplings for each $M$. It is evident that as the ensemble size $M$ increases, both RMSE and standard deviation tend to converge. Larger ensemble sizes exhibit reduced deviation, indicating enhanced stability. This observation suggests that, on average, larger $M$ yields more robust performance, as the influence of poor individual models is mitigated through ensemble averaging.

\begin{figure}[htbp]
    \centering
    \includegraphics[width=0.48\textwidth]{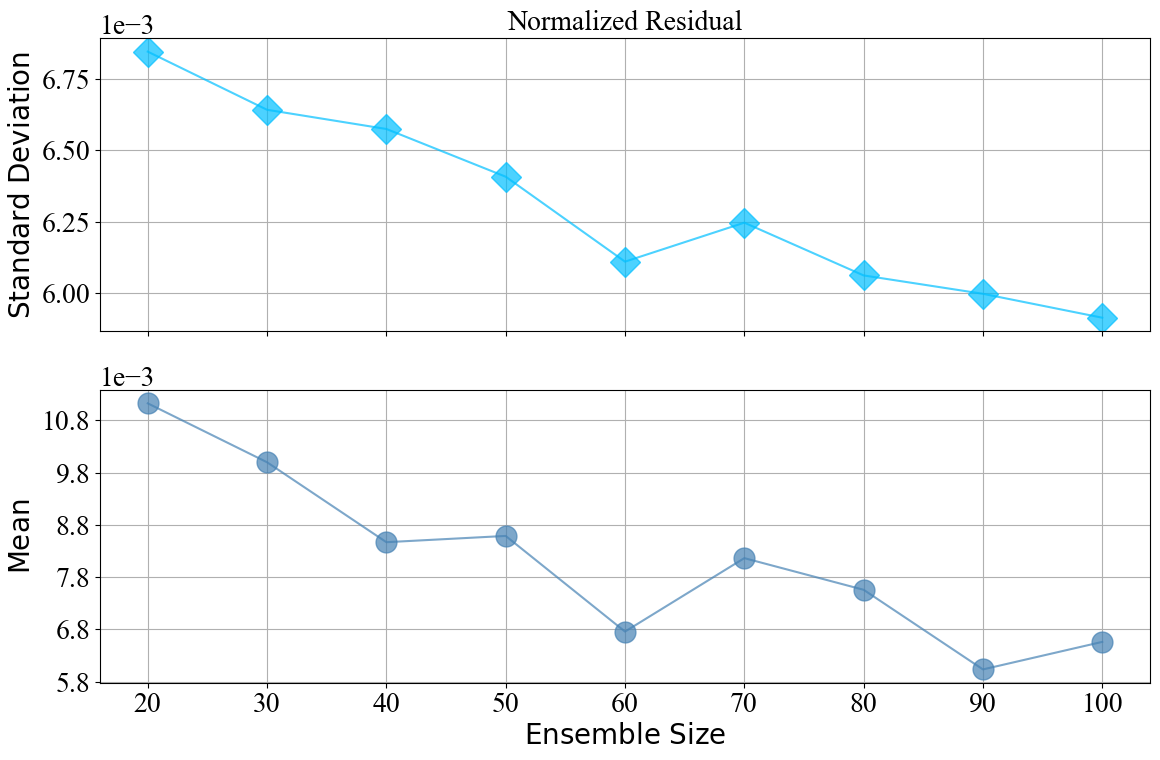}
    \caption{Standard Deviation ({\bf top}) and mean ({\bf bottom}) of the normalized residual $\hat{r}_0$ under various ensemble sizes $M$ of GAN with 51k parameters and batch size 102k.}
    \label{fig:ensemble_100}
\end{figure}

To gain insight into the advantages offered by ensemble methods, we expanded the ensemble size of largest model to 100. In Fig.~\ref{fig:ensemble_100}, we demonstrate that as $M$ increases, the residual decreases along with a reduction in the standard deviation.

In summary, ensemble methods applied to GANs with increased parameters and augmented datasets consistently exhibit superior performance, effectively addressing the critical necessity of scaling our GAN.

\subsection{Results from running the Distributed Training on Polaris}
 
\subsubsection{Time and Analysis Rate}
\begin{figure}[htbp]
    \centering
    \includegraphics[width=0.47\textwidth]{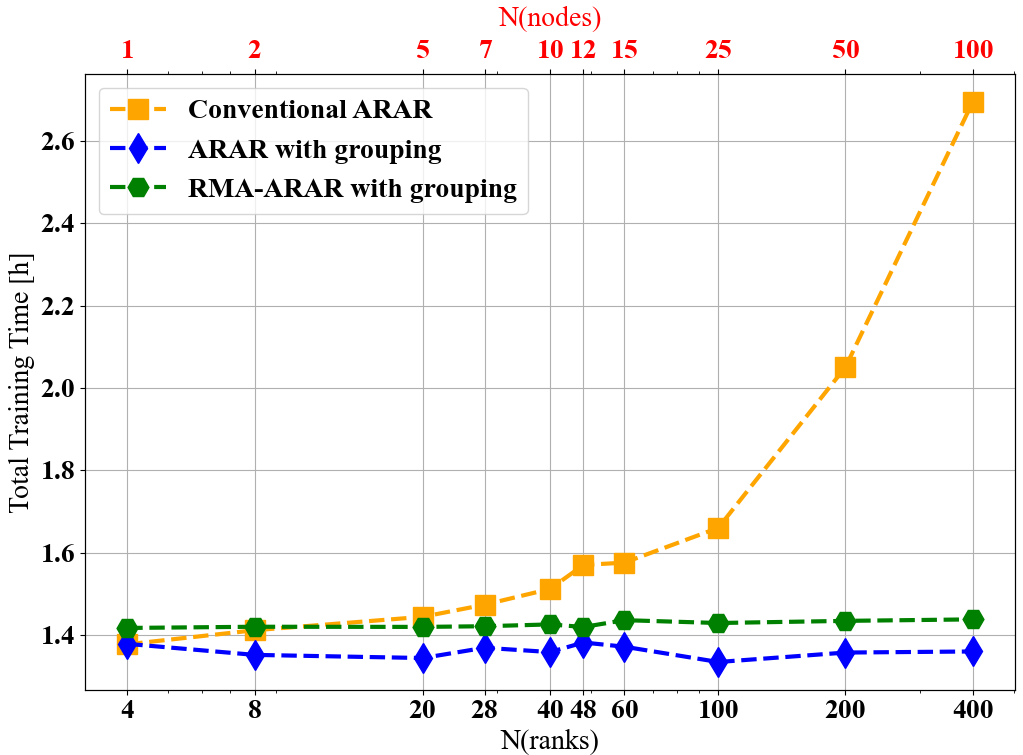}
    \caption{Total training time as a function of the number of ranks used to train the distributed GAN on Polaris. The top x-axis (red) represents the number of nodes which corresponds to the number of used ranks. Both x-axes are in logarithmic scale for proper visualization.}
    \label{fig:total_time_dist}
\end{figure}

\begin{figure}[htbp]
    \centering
    \includegraphics[width=0.47\textwidth]{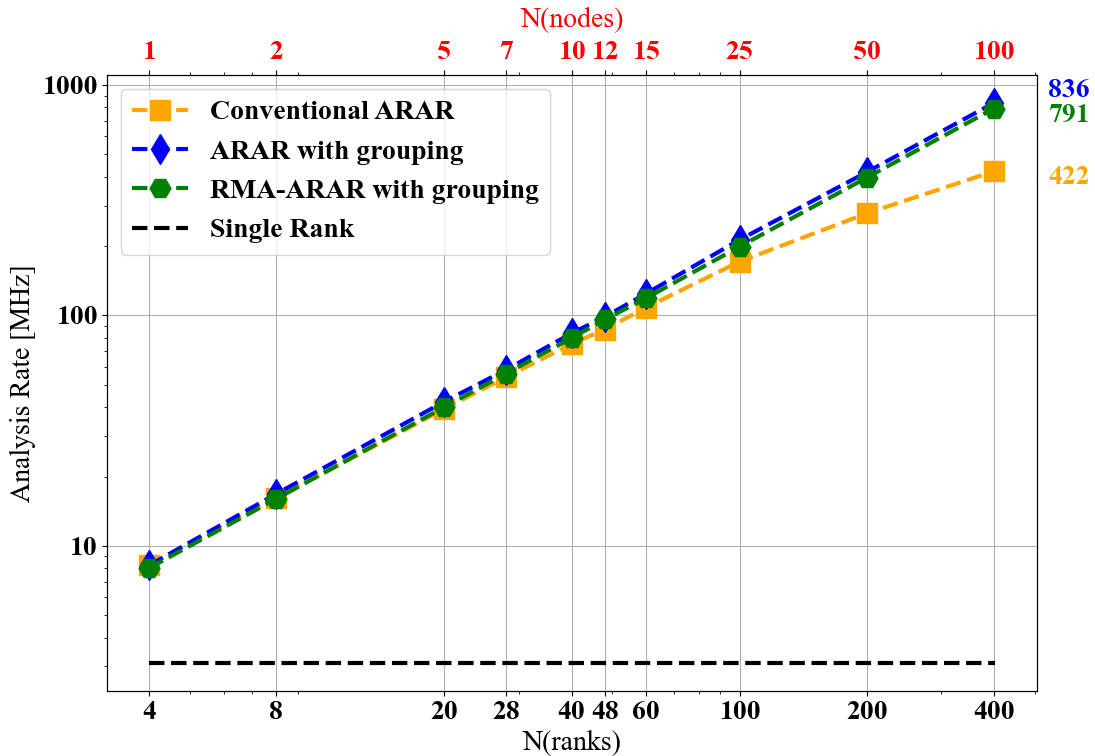}
    \caption{Plot of (\ref{def_ana_power}) as a function of the number of ranks used to train the distributed GAN. The horizontal dashed line represent the analysis rate for a single GAN, trained on one single GPU. The top x-axis (red) represents the number of nodes which corresponds to the number of used ranks. All axes use a logarithmic scale for proper visualization. The three numbers in the top right corner represent the analysis rate for N(ranks) = 400.}
    \label{fig:ana_power}
\end{figure}

Fig.~\ref{fig:total_time_dist} shows the total training time of the GAN workflow as a function of the number of ranks on Polaris. The corresponding number of compute nodes is represented by the top axis in Fig.~\ref{fig:total_time_dist}. One notices immediately that the training time for the conventional ARAR (Asynchronous Ring-All-Reduce) increases nearly linearly (the x-axis is in a logarithmic scale) w.r.t. the number of ranks involved. 
The ARAR and RMA-ARAR analyses (both with grouping) on the other hand, show nearly no dependency with respect to the number of ranks involved. Knowing the time dependencies allows us to formulate the analysis rate:
\begin{equation}
    \label{def_ana_power}
    \text{Analysis Rate} = \frac{N(\text{ranks}) \cdot N_{disc} \cdot N_{epochs}}{\text{Total Training Time}}
\end{equation}
where $N_{disc} = 1024,000$ is the number of discriminator samples (i.e. the discriminator batch size) and $N_{epochs}=100\,\rm{k}$ the number of training epochs. The analysis rate simply indicates how many events are analyzed by the GAN  within the given training time interval. A visualization of \eqref{def_ana_power} is shown in Fig.~\ref{fig:ana_power}. 
The analysis rates are similar for all methods for  $N(ranks)\lesssim 28$. After that the conventional ARAR method starts to saturate, whereas the RMA-ARAR and ARAR method (both with grouping) increase linearly. The gain in the analysis rate for the conventional ARAR is $\approx 40$ when going from 4 to 400 GPUs. The grouping mechanism used in this work allows doubling this gain.

\subsubsection{Convergence}
We will shift our focus to ensemble analyses, as they allow us to address the uncertainty of a given model. From now on, if not explicitly mentioned otherwise, the notation RMA-ARAR / ARAR implies the usage of the grouping mechanism presented in section~\ref{sec_grouping}. Furthermore, we use the notation (RMA-)ARAR to refer to either of the two distributed training methods.

Unlike the ensemble analysis discussed in the previous sections, the (RMA-)ARAR analysis does not utilize the entire input data set. For each rank, a random sub-sample of the input data is drawn and then used for training the GAN (see section~\ref{sec_distributed} for more details). The sub-sample size in this study corresponds to $50\%$ of the input data.
\begin{figure}[htbp]
    \centering
    \includegraphics[width=0.47\textwidth]{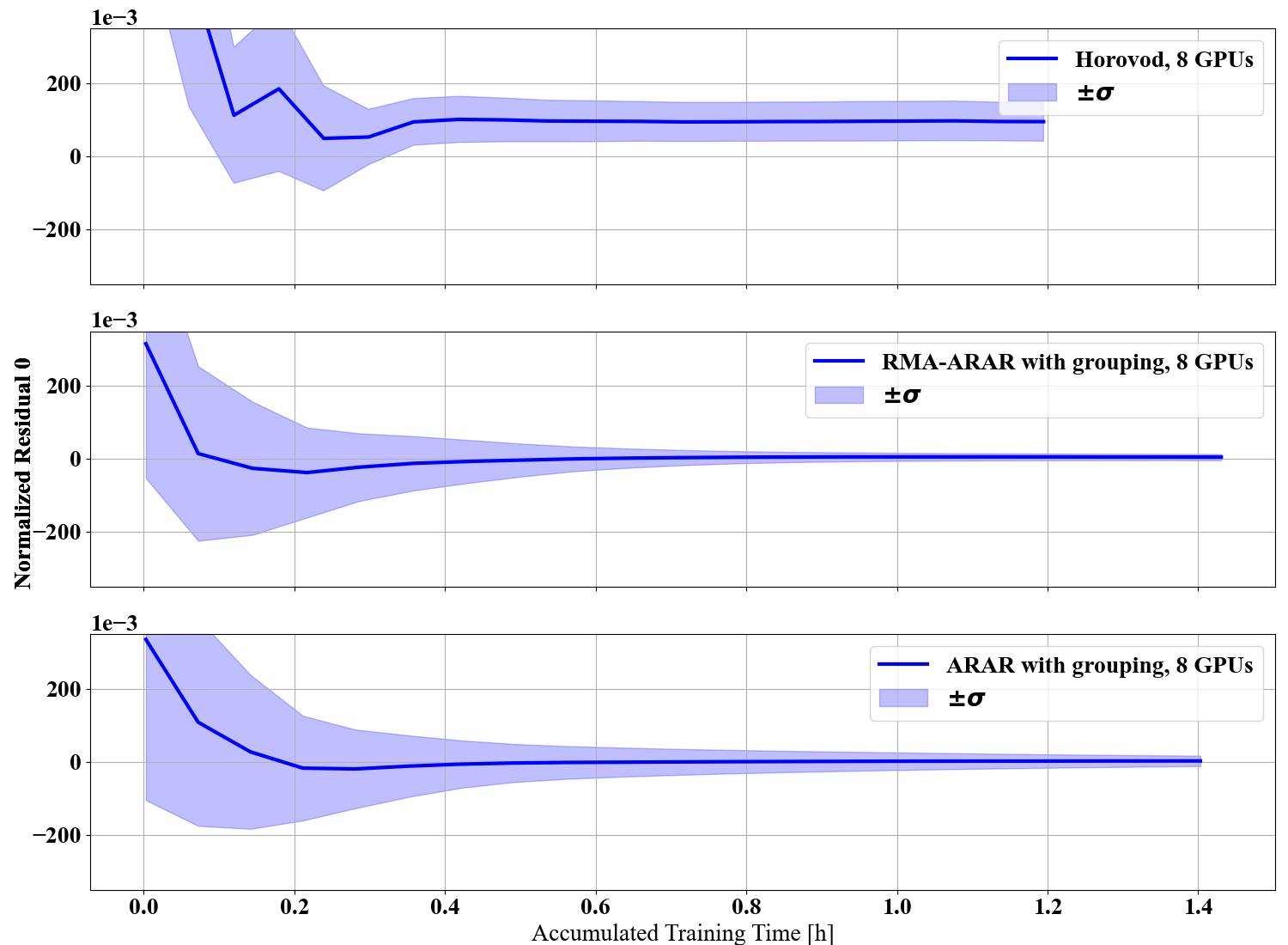}
    \caption{Normalized residual $\hat{r}_0$ from (~\ref{def_param_res}) as a function of the accumulated GAN training time. Each panel represents the response of an ensemble with 20 GAN generators. All GANs have been trained on 8 GPUs on Polaris. The shaded, blue areas in each panel correspond to a $1\sigma$ region around the ensemble predictions (solid, blue lines).}
    \label{fig:comp_res0_hvd}
\end{figure}

All results reported in this section are obtained from a post-training analysis. The GANs in each ensemble were trained and the states of the generator networks were stored together with a time stamp. These checkpoints were taken at the first epoch and every other $5\,\rm{k}$ epochs (resulting in 21 generator checkpoints). In combination with the time stamps, the checkpoints allow determining the convergence as a function of time which we will discuss in this section. 

Fig.~\ref{fig:comp_res0_hvd} summarizes the ensemble predictions for parameter 0. Shown is the normalized residual from \eqref{def_param_res} as a function of the accumulated training time. The first panel represents an analysis that was carried out with horovod, where every GAN within the ensemble was trained on 8 GPUs. Each horovod rank had access to the full input data size. The second to third panel in Fig.~\ref{fig:comp_res0_hvd}) summarize the normalized residuals from the (RMA-)ARAR ensemble analyses. The horovod analysis finished about $12\,\rm{min}$ earlier than the other methods. The corresponding convergence quality however is inferior to the (RMA-)ARAR analysis. Similar conclusions can be drawn from the remaining parameter residuals that are summarized in Tab.~\ref{tab:r_summary}. We also report the results (see fifth column) from running the distributed training via the conventional ARAR method. The results are consistent with those from the (RMA-)ARAR analysis. We repeated the above comparison by using 20 instead of 8 GPUs and obtained results similar to those shown in Tab.~\ref{tab:r_summary}. It is worth noting that previous horovod scaling studies~\cite{wu2018performance,10.1145/3337821.3337905} observed a decline in both accuracy and convergence with the addition of more ranks.
\begin{table}[htbp]
    \centering
    \caption{Normalized parameter residuals $\hat{r}_i$, together with a $1\sigma$ uncertainty, obtained from an ensemble analysis with horovod (second column), (RMA-)ARAR (third to fourth column) and conventional ARAR (fifth column). All numbers displayed here correspond to the last training time step, i.e. when the ensemble analysis has finished. Each analysis utilized 8 GPUs.}
    \begin{tabular}{c||c|c|c|c}
  Residual $[10^{-3}]$ & hvd & RMA-ARAR & ARAR & Conv. ARAR\\
  \hline
  \hline
  $\hat{r}_0$ & $95\pm 53$ & $5 \pm 9$ & $3\pm 14$ & $2 \pm 9$\\

  $\hat{r}_1$ & $94\pm 54$ & $6 \pm 14$ & $8 \pm 12$ & $3 \pm 13$\\

  $\hat{r}_2$ & $26 \pm 17$ & $1\pm 10$ & $0\pm 16$ & $0 \pm 9$\\
 
  $\hat{r}_3$ & $212 \pm 128$ & $24 \pm 21$ & $20\pm 19$ & $26 \pm 18$\\
 
  $\hat{r}_4$ & $138 \pm 85$ & $17 \pm 22$ & $14\pm 23$ & $18 \pm 20$\\
 
  $\hat{r}_5$ & $99 \pm 60$ & $11 \pm 8$ & $9\pm 9$ & $11 \pm 7$\\
 
    \end{tabular}
    \label{tab:r_summary}
\end{table}

\subsubsection{Comparison to single GPU Analysis}
Lastly, we would like to compare our method to an ensemble analysis that uses one GPU per GAN workflow. It should be noted that having a single GPU analysis as a reference is a luxury. Future analyses on actual measurements and using a more complex pipeline $f(\hat{\bm x}({\bm p}))$ will not allow running SAGIPS on a single GPU. 

Fig.~\ref{fig:ana_power} shows that adding more GPUs to the (RMA-)ARAR analysis, while keeping the discriminator batch size constant, increases the analysis rate (i.e. the number of events processed per second). Another way to look at this is to keep the analysis rate constant and check how the total training time varies, if more GPU resources are added to the (RMA-)ARAR analysis. In order to test this, the number of predicted parameter samples (see Tab.~\ref{tab:gen_settings}) was changed from $1024$ to:
\begin{equation}
    \label{split_batch}
    \#\text{ Predicted Parameter Samples} = \Big\lfloor \frac{1024}{\text{N(ranks)}} \Big\rfloor
\end{equation}
All other settings in Tab.~\ref{tab:gen_settings} remained the same which means that the discriminator batch size decreases with $1/\text{N(ranks)}$. We did explore the option to scale the generator learning rate w.r.t the number of ranks, but did not observe an improvement over the default settings reported in Section~\ref{sec_gan_workflow_settings}. We therefore kept the generator learning rate constant. 

\begin{figure}[htbp]
   \centering
   \includegraphics[width=0.47\textwidth]{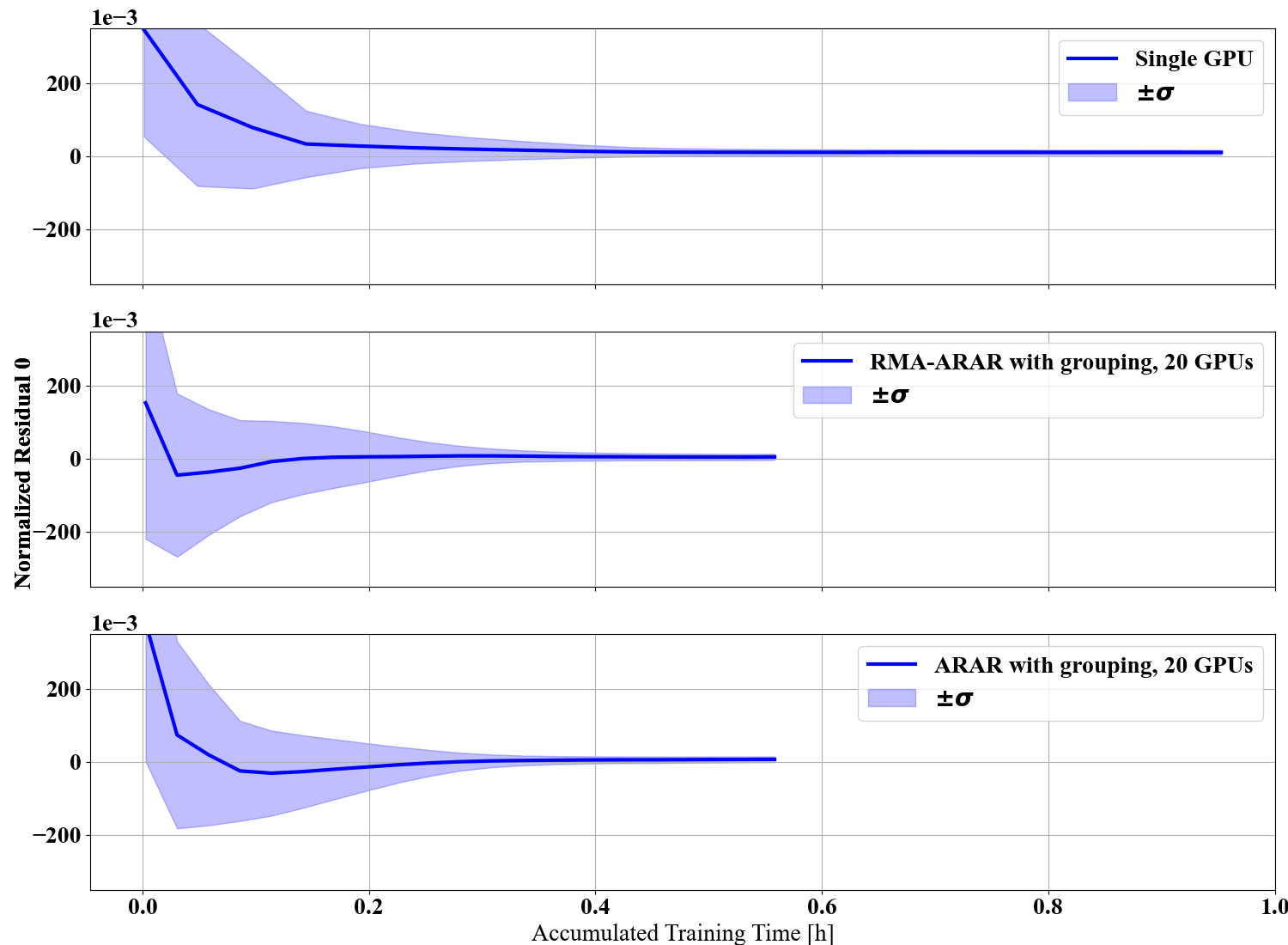}
   \caption{Normalized residual $\hat{r}_0$ from (\ref{def_param_res}) as a function of the accumulated GAN training time. The top panel refers to the ensemble analysis carried out on GPU, whereas the second / third panel from the top represent the RMA-ARAR / ARAR ensemble analysis with 20 GPUs. The solid blue curves represent the ensemble predictions, while the blue shaded areas indicate a $1\sigma$ uncertainty.}
   \label{fig:comp_res0_split_batch}
\end{figure}

\begin{figure}[htbp]
    \centering
    \includegraphics[width=0.47\textwidth]{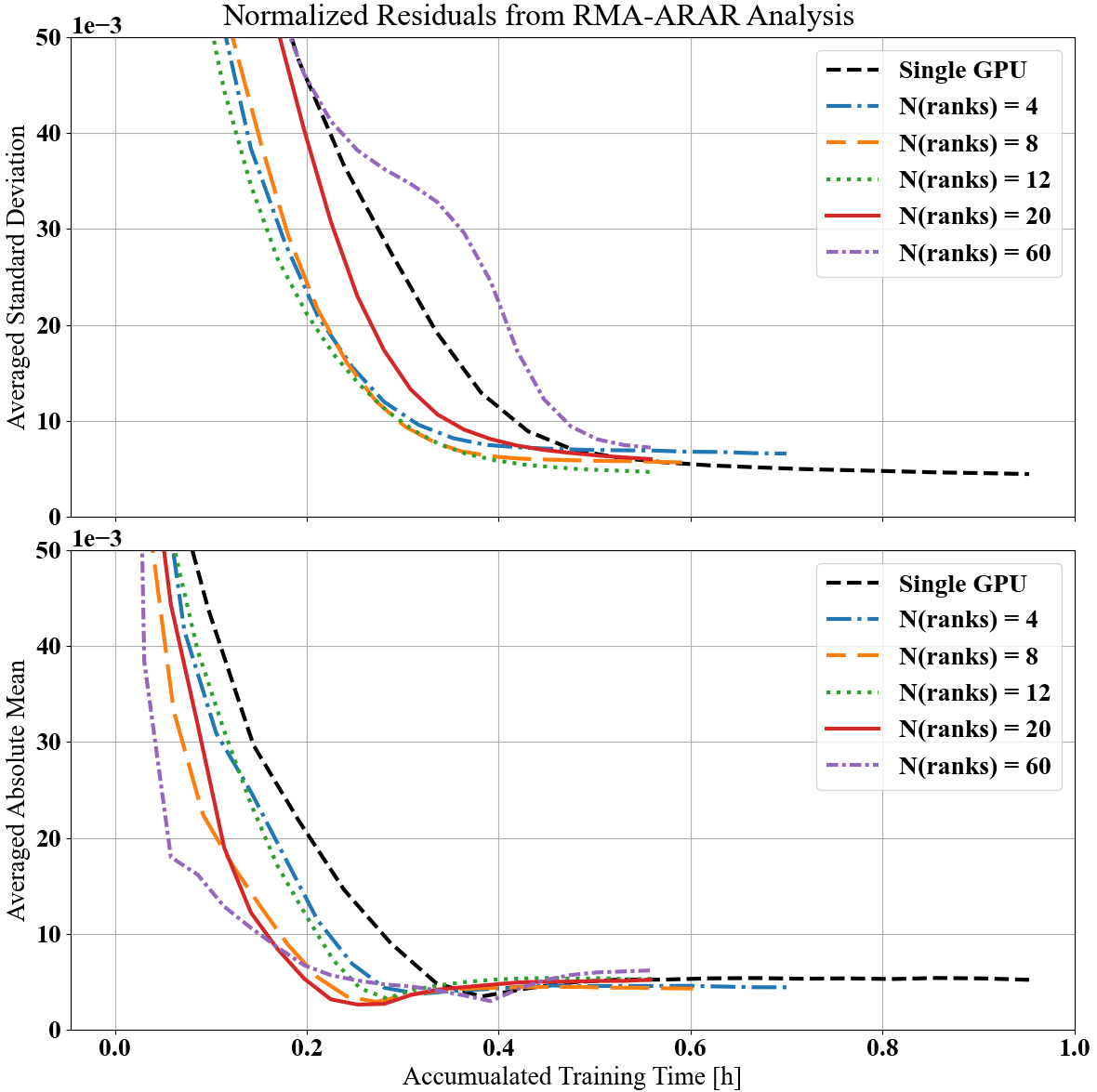}
    \caption{Standard Deviation ({\bf top}) and mean ({\bf bottom}) of the normalized residuals as a function of the GAN training time. Each quantity represents the average over all parameters and is scaled by a factor 1000 for convenience. The black, dashed line represents the results deduced from the single GPU analysis. The remaining curves corresponds to the results obtained from the RMA-ARAR training, using 2,4,8, 20 and 60 GPUs.}
    \label{fig:conv_rma_arar}
\end{figure}

\begin{figure}[htbp]
    \centering
    \includegraphics[width=0.47\textwidth]{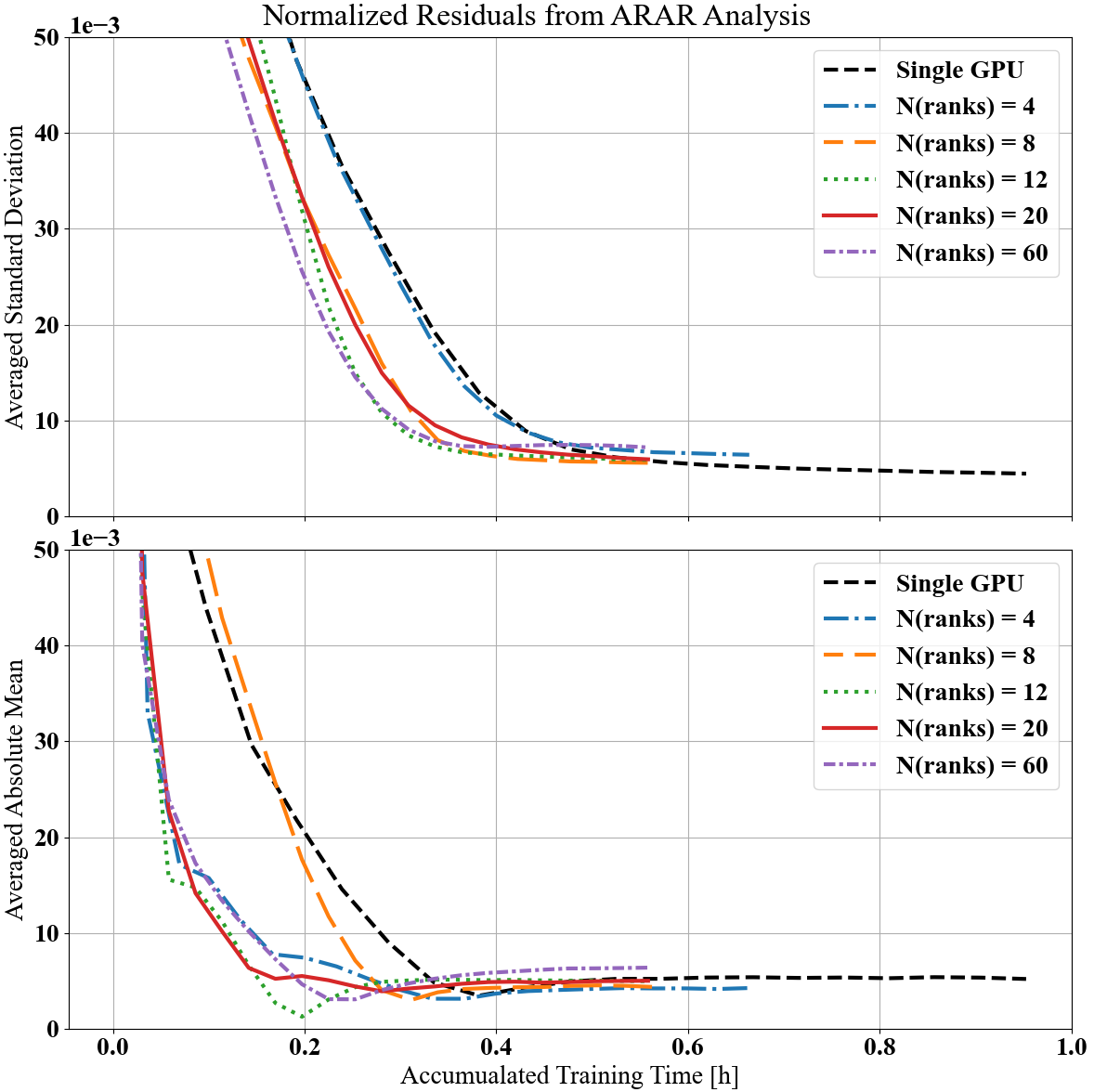}
    \caption{Standard Deviation ({\bf top}) and mean ({\bf bottom}) of the normalized residuals as a function of the GAN training time. Each quantity represents the average over all parameters and is scaled by a factor 1000 for convenience. The black, dashed line represents the results deduced from the single GPU analysis. The remaining curves corresponds to the results obtained from the ARAR training, using 2,4,8, 20 and 60 GPUs.}
    \label{fig:conv_arar}
\end{figure}

Fig.~\ref{fig:comp_res0_split_batch} shows the outcome of using \eqref{split_batch} in an (RMA-)ARAR ensemble analysis. One immediately notices that the total training time is, compared to the single GPU one, noticeably reduced. The convergence quality however is consistent between the single- and multi-GPU analyses. 
This becomes more clear when inspecting Fig.~\ref{fig:conv_rma_arar} and~\ref{fig:conv_arar}. The results shown there indicate that the GAN learns faster if multiple GPUs are utilized and moreover suggest that the distributed training might be terminated earlier at $\approx 0.4\,\rm{h}$. We determined this time by inspecting the intersection point between the single- and multi-GPU curves in the bottom panels of Fig.~\ref{fig:conv_rma_arar} and~\ref{fig:conv_arar}. The scaling behavior depicted in both figures is not intuitive. Our explanations for this are two-fold: Firstly, the generator predictions in our workflow are not directly passed to the discriminator, but rather sent through a pipeline first. Depending on the pipeline, the generated and input observables might show good agreement, while the predicted parameters have not yet fully converged to the optimal solution. Efforts to investigate this in more detail are currently ongoing. Secondly, the discriminator batch size decreases w.r.t. ~\eqref{split_batch}, which ultimately affects the generator training and therefore has an impact on the parameter predictions. 

From the experiments discussed here, we conclude that the (RMA-)ARAR analysis can be run with less data (per rank) and a smaller discriminator batch size without loosing convergence quality (compared to a single GPU analysis). A smaller discriminator batch size means a reduced utilization of the pipeline. Depending on the complexity of the pipeline, this will have a significant impact on the overall GAN training time. In our case, using a very simple pipeline, we saved $\approx 48\,\rm{min}$ when going from a discriminator batch size of 102,400 to 5,100 . We would like to emphasize that, even though we used it for benchmarking, a single GPU ensemble analysis will not be a suitable option for running our GAN workflow on real physics data. The anticipated data volume as well as the complexity of the pipeline will not support a single GPU utilization. 

\section{Summary and Outlook}
In this work, we examined various mechanisms to execute a generative, inverse problem solving algorithm (SAGIPS) across multiple compute nodes. 
Key challenges were posed by the GAN nature of the algorithm as well as the SAGIPS pipeline which may introduce delays between ranks in a multi-GPU analysis. 
We identified two methods that are based on data parallel training with overlap: (i) ARAR with grouping and (ii) RMA-ARAR with grouping. 
For the work presented in this paper, the GAN generator is trained in an asynchronous ring-all-reduce fashion across multiple ranks, while each rank has its own discriminator network. 
We showed that by grouping GPUs per node the total training time is reduces and scales nearly linearly. 
Moreover, we found that utilizing RMA as a mean to transfer gradients was a viable option for the asynchronous data parallel training. 
Future studies entail the use of SAGIPS together with (RMA-)ARAR on more complex and resource intensive pipelines.
We plan to test methods such as tensor gradient fusion or splitting gradient tensors into smaller tensor packages, as an addition to the asynchronous ring-all-reduce mechanism. Furthermore, we would like to explore the double binary tree~\cite{binarytree2013} method for gradient transfer between GPUs. 
\section{Acknowledgements}
This work is supported by the Scientific Discovery through Advanced
Computing (SciDAC) program via the Office of Nuclear Physics and Office of
Advanced Scientific Computing Research in the Office of Science at the U.S.
Department of Energy under contracts DE-AC02-06CH11357, DE-AC05-06OR23177,
and DE-SC0023472, in collaboration with Argonne National Laboratory,
Jefferson Lab, National Renewable Energy Laboratory, Old Dominion
University, Ohio State University, and Virginia Tech.
This research used resources of the Argonne Leadership Computing Facility, a U.S. Department of Energy (DOE) Office of Science user facility at Argonne National Laboratory and is based on research supported by the U.S. DOE Office of Science-Advanced Scientific Computing Research Program, under Contract No. DE-AC02-06CH11357.

\bibliographystyle{IEEEtran}
\bibliography{reference}

\end{document}